\documentclass[aps,pra,twocolumn,showpacs,superscriptaddress]{revtex4}

\usepackage{graphicx}

\def\figwidth{8.5cm}

\begin{document}
\newcommand{\beq}{\begin{equation}}
\newcommand{\eeq}{\end{equation}}
\newcommand{\barr}{\begin{eqnarray}}
\newcommand{\earr}{\end{eqnarray}}

\newcommand{\andy}[1]{ }

\def\coltwovector#1#2{\left({#1\atop#2}\right)}
\def\upp{\coltwovector10}
\def\downn{\coltwovector01}
\def\bra#1{\langle #1 |}
\def\ket#1{| #1 \rangle}
\def\Ord{\mbox{O}}
%
%
\def\ask{\marginpar{?? ask:  \hfill}}
\def\fin{\marginpar{fill in ... \hfill}}
\def\spnote{\marginpar{note (SP) \hfill}}
\def\Olomouc{\marginpar{note (JR,ZH) \hfill}}
\def\pfnote{\marginpar{note (PF) \hfill}}
\def\check{\marginpar{check \hfill}}
\def\discuss{\marginpar{discuss \hfill}}
%
%

\title{Quantum Zeno tomography}

\author{P. Facchi}
\affiliation{Dipartimento di Fisica, Universit\`a di Bari  I-70126 Bari,
Italy}
\affiliation{Istituto Nazionale di Fisica Nucleare, Sezione di Bari,
I-70126 Bari, Italy }
\author{Z. Hradil}
\affiliation{Department of Optics, Palack\'{y} University,
17. listopadu 50, 772~00 Olomouc, Czech Republic }
\author{G. Krenn}
\affiliation{Dipartimento di Fisica, Universit\`a di Bari  I-70126 Bari,
Italy}
\author{S. Pascazio}
\affiliation{Dipartimento di Fisica, Universit\`a di Bari  I-70126 Bari,
Italy}
\affiliation{Istituto Nazionale di Fisica Nucleare, Sezione di Bari,
I-70126 Bari, Italy }
\author{J. \v{R}eh\'{a}\v{c}ek}
\affiliation{Department of Optics, Palack\'{y} University,
17. listopadu 50, 772~00 Olomouc, Czech Republic }

\date{\today}

\begin{abstract}
We show that the resolution ``per absorbed particle'' of standard
absorption tomography can be outperformed by a simple
interferometric setup, provided that the different levels of
``gray'' in the sample are not uniformly distributed. The
technique hinges upon the quantum Zeno effect and has been tested
in numerical simulations. The scheme we propose could be
implemented in experiments with UV-light, neutrons or X-rays.
\end{abstract}

\pacs{03.65.Xp}

\maketitle

\section{Introduction}
\label{sec-intr}\andy{sec-intr}

Absorption tomography is an important experimental technique
revealing the internal structure of material bodies. By measuring
the attenuation of a beam of particles passing through a sample
one infers the absorption coefficient (density) of the sample in
the beam section. The possibility of distinguishing two slightly
different densities of the material is often of vital importance.
Under otherwise ideal conditions the shot noise associated with
the discrete character of the illuminating beam sets an upper
limit to the resolution of absorption tomography: for instance,
the shadow cast by a brain tumor might become totally lost in the
noisy data. One possibility to overcome the fluctuations is to
increase the intensity of the beam. However, in many situations,
like in medicine for example, the intensity of the illuminating
beam cannot be made arbitrarily high due to the damage provoked
by the absorbed radiation.

A significant step towards an ``absorption-free tomography" came
from quantum theory. It was demonstrated, both theoretically
\cite{inter-free-theor1,inter-free-theor2} and experimentally
\cite{inter-free-exper}, that \textit{totally} transmitting and
absorbing bodies can be distinguished without absorbing any
particles, by using an interferometric setup. This idea is in
fact a clever implementation of the quantum Zeno effect
\cite{Misra} and hinges upon the notion of ``interaction-free"
measurement \cite{Dicke}. A classical measuring apparatus (here
the black sample), placed in one arm of the interferometer,
projects the illuminating particle into the other arm, destroying
interference, freezing the evolution and forcing the particle to
exit through a different channel from that it would have chosen
had both arms been transparent (white sample).

In practical applications, however, samples are normally neither
black nor white: they are gray. In this paper we endeavor to
understand whether application of the quantum Zeno effect, which
turns out to be ideal for discriminating black and white, might be
advantageous also for the more practical task of discriminating
two gray bodies with different transmission coefficients. More
specifically, we ask: is it possible by quantum Zeno effect to
reduce the number of absorbed particles while preserving the
resolution? We show that this is indeed possible. Closely related
questions have been recently investigated by other authors
\cite{KSS,MM}. Our conclusions are somewhat optimistic: we show
that standard absorption tomography can be outperformed by a Zeno
setup, provided that the frequency of occurrences of the different
levels of ``gray" in the sample is not uniform. More to this, the
Zeno setup, unlike the standard one, is endowed with two detection
channels: as we shall see, this feature, if properly exploited,
leads to even better performances in the Zeno case.

\section{Quantum Zeno effect in a Mach-Zehnder interferometer}
\label{sec-MZI}\andy{sec-MZI}

\begin{figure}
\includegraphics[width=\figwidth]{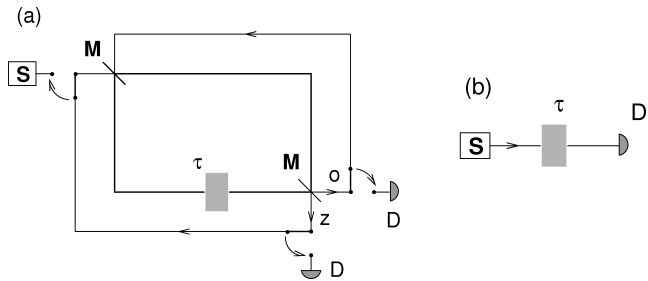}
\caption{\label{fig:outline} a) Scheme
of the Zeno interferometric setup. b) Standard transmission
experiment. S -- source; M -- semi-transparent mirror; o --
orthogonal channel; z --Zeno channel; D -- detector.}
\end{figure}

We introduce notation and sketch the fundamental features of the
quantum Zeno effect. We consider the Mach-Zehnder interferometric
(MZI) scheme with feedback displayed in Fig.\ \ref{fig:outline}a.
A semitransparent object, whose transmission amplitude is $\tau$
(assumed real for simplicity) is placed in the lower arm of the
interferometer. The particle is initially injected from the left,
crosses the interferometer $L$ times and is finally detected by
one of two detectors. The two semitransparent mirrors M are
identical and their amplitude transmission and reflection
coefficients are
\andy{coeffs}
\beq
\label{eq:coeffs}
c \equiv \cos \theta_L, \quad s \equiv \sin
\theta_L \quad (\theta_L=\pi/4L),
\eeq
respectively. Notice that both coefficients depend on $L$, the
number of ``loops" in the MZI.

The incoming state of the particle (coming from the source at
initial time) is
\andy{instate}
\beq
\label{eq:instate}
\ket{\mathrm{in}} = \upp
\eeq
and we call ``Zeno" and ``orthogonal" channels the extraordinary
$\upp$ and ordinary $\downn$ channels of the MZI, respectively.
The total effect of the interferometer is
\andy{BAB}
\beq
V_\tau=BA_\tau B, \quad B=\left(
\begin{array}{cc}
c & -s\\ s & c
\end{array}
\right),
\quad A_\tau=\left(
\begin{array}{cc}
1 & 0\\ 0 & \tau
\end{array}
\right).
\label{eq:BAB}
\eeq
In general,
\andy{Bdef}
\beq
B=\exp(-i\theta_L \sigma_2), \quad BB^\dagger=B^\dagger B=1,
\eeq
where $\sigma_2$ is the second Pauli matrix, while $A_\tau$ is not
unitary (if $\tau<1$ there is a probability loss). The final
state, after the particle has gone through $L$ loops, reads
\andy{fin}
\beq
\ket{\mathrm{out}} = V_\tau^L \ket{\mathrm{in}}= (BA_\tau B)^L
\ket{\mathrm{in}}.
\label{eq:fin}
\eeq

\subsection{White sample}
\label{sec-white}\andy{sec-white}

The choice of the angle $\theta_L$ in (\ref{eq:coeffs}) is
motivated by our requirement that if $\tau=1$ (``white" sample,
i.e.\ no semitransparent object in the MZI) the particle ends up
in the ``orthogonal" channel:
\andy{V1}
\beq
V^L_{\tau=1} = B^{2L}= e^{-i2L\theta_L \sigma_2}=
e^{-i\pi\sigma_2/2}= -i \sigma_2,
\label{eq:V1}
\eeq
so that
\andy{finV1}
\beq
\ket{\mathrm{out}} = V^L_{\tau=1} \ket{\mathrm{in}} =
 \downn .
\label{eq:finV1}
\eeq
This is easy to understand: each loop ``rotates" the particle's
state by $2\theta_L=\pi/2L$ and after $L$ loops the final state is
``orthogonal" to the initial one (\ref{eq:instate}).

\subsection{Black sample}
\label{sec-black}\andy{sec-black}

Let us now look at the case $\tau=0$, corresponding to a
completely opaque (``black") object in the MZI. We obtain
\andy{V0}
\barr
V^L_{\tau=0} &=& B(A_0 B^2)^LB^{-1} \nonumber \\
&=& B \cos^L 2\theta_L\left(
\begin{array}{cc}
1 & -\tan 2\theta_L\\ 0 & 0
\end{array}
\right)B^{-1}\nonumber\\
&
\stackrel{L \to
\infty}{\longrightarrow} &
\left(
\begin{array}{cc}
1 & 0\\ 0 & 0
\end{array}
\right) \equiv {\cal V}_{\tau=0}.
\label{eq:V0}
\earr
This yields QZE:
\andy{outin}
\beq
\ket{\mathrm{out}} = {\cal V}_{\tau=0} \ket{\mathrm{in}} =
\ket{\mathrm{in}} = \upp.
\eeq
In the infinite $L$ limit the initial state is ``frozen" and the
particle ends up in the Zeno channel.

\subsection{Gray sample}
\label{sec-gray}\andy{sec-gray}

What happens if $0<\tau<1$? We easily get
\andy{Vtau}
\beq
V_\tau =
\left(
\begin{array}{cc}
(1+\tau)c^2-\tau & -sc(1+\tau) \\ sc(1+\tau) & \tau -(1+\tau)s^2
\end{array}
\right).
\label{eq:Vtau}
\eeq
The computation of $V^L_\tau$ is straightforward but lengthy and
yields a final expression which is elementary but complicated.
However, we are mainly interested in the large-$L$ limit, that for
\andy{tauzL}
\beq
\tau<\tau_L^{\mathrm{Z}} \equiv (1-\sin \pi/2L)/(1+\sin \pi/2L)
\label{eq:tauzL}
\eeq
reads \footnote{It is worth stressing that when
\(\tau>\tau_L^{\mathrm{Z}}\) the asymptotic expansion has a completely
different form and tends to $-i\sigma_2$ when $L\to
\infty$. Upon crossing the threshold $\tau=\tau_L^{\mathrm{Z}}$, the
apparatus starts operating in the Zeno regime. We also note that
$\tau_L^{\mathrm{Z}}$ in (\ref{eq:tauzL}) is bounded from below
according to the simple expression $\tau_L^{\mathrm{Z}}>1-\pi/L$.
Alternatively, one can rephrase these conditions in terms of $L$:
the asymptotic expansion (\ref{eq:eigen}) is then valid for $L >
L^{\mathrm{Z}}_\tau =
\pi/2 \arcsin [(1-\tau)/(1+\tau)]$.}
\andy{eigen}
\barr
V_\tau^L = \left(
\begin{array}{cc}
1-\frac{\pi^2}{8L} \frac{1+\tau}{1-\tau} & \Ord(L^{-1})\\
\Ord(L^{-1}) & \tau^L [1+\Ord(L^{-1})]
\end{array}
\right)+ \Ord(L^{-2}).
\label{eq:eigen}
\earr
This is an interesting result: indeed
\andy{oper}
\beq
{\cal V}_{\tau} \equiv \lim_{L \to\infty} V_\tau^L =
\left(
\begin{array}{cc}
1 & 0 \\ 0 & 0
\end{array}
\right),
\quad 0 \leq \tau <1
\label{eq:oper}
\eeq
analogously to (\ref{eq:V0}). This shows that even for a
semitransparent object, with transmission coefficient $\tau \neq
1$, a \textit{bona fide} QZE takes place and the particle ends up
in the Zeno channel with probability one:
\andy{stato}
\beq
\ket{\mathrm{out}} = {\cal V}_{\tau} \ket{\mathrm{in}} = \ket{\mathrm{in}}
, \qquad \tau\neq 1.
\label{eq:stato}
\eeq

\section{Distinguishing different shades of gray}
\label{sec-graydis}\andy{sec-graydis}

A question arises \cite{MM}: is it possible to distinguish
different values of $\tau$ (different ``shades" or ``levels" of
gray) by the technique outlined above? This is not a simple task,
for after a large number of loops $L$ the particle ends up in the
orthogonal channel only if $\tau=1$ [see Eq.\ (\ref{eq:finV1})];
by contrast, for any value of $\tau \neq 1$, the particle ends up
in the Zeno channel [see (\ref{eq:stato})] \textit{irrespectively}
of the particular value of $\tau$. However, the asymptotic
correction in the $(1,1)$ element of $V_{\tau}^L$ in
(\ref{eq:eigen}) is $\tau$ dependent: the details of the
convergence to the limit (\ref{eq:oper})-(\ref{eq:stato}) depend
on the grayness of the sample. By exploiting this feature, we
shall now show that it is indeed possible to resolve different
gray levels by QZE, within a given statistical accuracy.

We start by observing that if one performs a standard transmission
experiment, by shining a particle beam on a semitransparent object
in order to measure the transmission coefficient $\tau$, see Fig.\
\ref{fig:outline}b, the detection and absorption probabilities read
\andy{pad}
\beq
\label{eq:pad}
p'_d(\tau) = \tau^2, \qquad p'_a(\tau)= 1-\tau^2.
\eeq
The statistics is binomial.

On the other hand, if one uses the Zeno configuration sketched in
Fig.\ \ref{fig:outline}a, the final state of the particle after
$L$ loops in the MZI is, from (\ref{eq:eigen}),
\andy{amplu}
\begin{equation}
V_\tau^L \left(
\begin{array}{c}
1\\
0
\end{array}
\right) =\left(
\begin{array}{c}
u_z\\
u_o
\end{array}
\right) =
\left(
\begin{array}{c}
1-\frac{\pi^2}{8L} \frac{1+\tau}{1-\tau} + \Ord(L^{-2})\\
\Ord(L^{-1})
\end{array}
\right),
\label{eq:amplu}
\end{equation}
where $u_z$ and $u_o$ are the amplitudes
in the Zeno and orthogonal
channels, respectively. Both these quantities are real. Let
$(0\leq \tau < 1)$
\andy{pzoa}
\barr
p_z(\tau)&=& u_z^2=1-\frac{\pi^2}{4L}\frac{1+\tau}{1-\tau}
+\Ord(L^{-2}), \nonumber\\
\label{eq:pzoa}
p_o(\tau)&=& u_o^2=\Ord(L^{-2}),  \\
p_a(\tau)&=&1-p_z(\tau)-p_o(\tau)
=\frac{\pi^2}{4L}\frac{1+\tau}{1-\tau} +\Ord(L^{-2}) \nonumber
\earr
be the probabilities that the particle is detected in the Zeno,
orthogonal channel or is absorbed by the semitransparent object,
respectively. We assume that a fixed number of particles $N$ is
sent in the MZI during an experimental run. In this situation the
distribution of particles in the Zeno, orthogonal or absorption
channels follows a \textit{trinomial} statistics with
probabilities (\ref{eq:pzoa}).

It seems natural to think that, since by increasing the number of
loops $L$, $p_o$ vanishes much faster than $p_a$, for large $L$
the distribution is practically binomial with $p_a+p_z\approx1$
\footnote{More so, since $p_a$ is also small in this limit, the
detection statistics is almost Poissonian.}. However, as we shall
see, the presence of a small trinomial component, $p_o$, will play
an important role, enabling the Zeno method to perform much better
than the standard one.

\section{An intrinsic limit for binomial statistics: the Cram\'er-Rao bound}
\label{sec-CRbound}\andy{sec-CRbound}

We are now ready to discuss the possibility of a ``Zeno
tomography." The goal is to get information about the distribution
of the absorption coefficient in the sample, absorbing as few
particles as possible. We will accomplish this in two steps.
First, using estimation theory, we show that \textit{if one limits
one's attention to binomial statistics}, the Zeno estimation of
\textit{any} level of gray of one pixel (i.e.\ $\tau$ continuously
distributed between 0 and 1) \textit{cannot} perform better than
the standard method.  At best, both methods are equivalent. This
is bad news. However, the very proof of the above-mentioned
statement will show that there are two ways out: first, we will
see that Zeno performs better when one wants to distinguish two
levels of gray that are not equally populated in the sample (this
requires some prior knowledge about the distribution of grays in
the sample). This is good news, for it enables one to find a
method that in some cases works better than the standard one.
Second, one is led to think that the introduction and exploitation
of a trinomial statistics can enable the Zeno method to perform
better.

Let us start from the estimation of \textit{any} level of gray. In
this case one tries to estimate $\tau^2$ from the counted number
of particles, absorbing as few particles as possible for the
requested precision. To perform this task in an optimal way one
should find optimal estimators for each scheme. A lower bound on
the variance of an unbiased estimator $\hat{T}$ of the parameter
$T$ (here $T=\tau^2$) is the Cram\'er-Rao lower bound (CRLB)
\cite{Cramer},
\begin{equation}
\label{eq:crlb}
(\Delta \hat
T)^2\ge\frac{1}{F}\equiv\left\langle\left[\frac{\partial}{\partial
T}\ln p(n|T)\right]^2\right\rangle^{-1},
\end{equation}
where $F$ is the Fisher information, $p(n|T)$ the probability of
observing $n$ particles conditioned by the value $T$ of the
unknown parameter and $\langle\dots\rangle$ denotes ensemble
average with respect to $n$. The probability $p$ is binomially
distributed in both the standard case and the Zeno case, which
yields
\andy{nastnaze}
\barr
(\Delta \hat T_{\mathrm{st}})^2 &\ge& \frac{\tau^2(1-\tau^2)}{N},
\nonumber \\
(\Delta
\hat T_{\mathrm{Ze}})^2& \ge &
\frac{4\tau^2(1-\tau)^3(1+\tau)L}{\pi^2N}
\label{eq:nastnaze}
\earr
for standard and Zeno tomography, respectively, $N$ being the
(fixed) number of input particles in both cases. Expressing the
above inequalities (\ref{eq:nastnaze}) in terms of the number of
\textit{absorbed} particles $N_a^{\mathrm{(st)}}=N p'_a$ and
$N_a^{\mathrm{(Ze)}}=N p_a$, they both reduce to the same bound
\andy{crlb-sta}
\begin{equation}
\label{eq:crlbsta}
\Delta \hat T_{\mathrm{st,Ze}}^{\,\mathrm{opt}}
\ge\frac{\tau(1-\tau^2)}{\sqrt{N_a^{\mathrm{(st),(Ze)}}}},
\end{equation}
showing that the CRLB's for standard and Zeno tomography are the
same, \textit{given} the number of absorbed particles. Further, it
is trivial to show that the unbiased estimator given by the
relative frequency of transmitted particles, $\hat
T_{\mathrm{st}}=n_t/N$, saturates the CRLB (\ref{eq:crlbsta}).
Hence,
\textit{if one neglects the output of the ordinary channel $p_o$}
and considers the statistics (\ref{eq:pzoa}) \textit{practically}
binomial, the Zeno estimation can be at most as good as the
standard one: it cannot be better.

\section{Two statistical protocols}
\label{sec-statprot}\andy{sec-statprot}

\subsection{Binomial (single-channel) protocol}

In spite of the conclusions of Sec.\ \ref{sec-CRbound}, we will
now construct a protocol and show that by QZE one can achieve a
resolution that is \textit{superior} to the ``ordinary" resolution
obtained in a standard transmission experiment. Notice that
$p_a(\tau)$ in (\ref{eq:pzoa}), unlike $p'_a(\tau)$ in
(\ref{eq:pad}), is an \textit{increasing} function of $\tau$.
Therefore, with respect to absorbed particles, the Zeno
tomographic image (for sufficiently large $L$) yields a kind of
negative of the standard absorption tomographic image. This can be
given a rather intuitive explanation: indeed, the absorption
probability in\ (\ref{eq:pzoa}) reduces to the same form as the
standard one\ (\ref{eq:pad}), i.e.\
\andy{padz}
\beq
\label{eq:padz}
p_a(\tau)= 1-\left(\tau^{\mathrm{Ze}}_{\mathrm{eff}}\right)^2,
\eeq
by introducing an ``effective" transmission coefficient
\andy{teff}
\beq \label{eq:teff}
\tau^{\mathrm{Ze}}_{\mathrm{eff}}=\sqrt{1-p_a}.
\eeq
For example, if we take $\tau_1=0.98$, $\tau_2=0.99$ and choose
$L=12000$, then, according to Eq.~(\ref{eq:pzoa}), we get
$\tau_{\mathrm{eff}1}^{\mathrm{Ze}}\approx 0.99$ and
$\tau_{\mathrm{eff}2}^{\mathrm{Ze}}\approx 0.98$. The two gray
levels are
\textit{interchanged} by the Zeno apparatus. If most of the sample
has transmission coefficient $\tau_2$ the absorbed energy is
reduced by using the Zeno setup.

A more precise comparison of the performances of the Zeno and
standard techniques can be given in the framework of decision
theory. For simplicity let us focus on distinguishing only two
gray levels $\tau_1$ and $\tau_2$ ($\tau_1<\tau_2$) corresponding
to hypotheses $H_1$ and $H_2$ that occur in the sample with
frequencies
\andy{freqa}
\beq \label{eq:freqa}
P_0(H_1)=\alpha, \quad P_0(H_2)=1-\alpha.
\eeq
With this simplification we lose no generality since
the tomography with $M$ gray levels can always be split
into a sequence of pairwise decisions between two adjacent
gray levels.

We will proceed in two steps. First we will assume that in the
Zeno configuration of Fig.\ \ref{fig:outline}a all output
particles are collected at a single detector. In other words, the
Zeno and orthogonal channels are considered as a single output and
the statistics (\ref{eq:pzoa}) is binomial ($p_z+p_o, p_a)$. Each
particle is then either absorbed or transmitted (and detected) by
the Zeno apparatus. Obviously, by merging the two output channels
together, some information about the sample is wasted. We know,
however, that this strategy will be optimal if the number of loops
$L$ is very large. Then there are almost no particles exiting via
the ordinary channel $n_o\approx 0$ which can then be safely
ignored. A better and more general strategy will be studied later.

Since both experiments obey the same (binomial) statistics we use
the notation of the Zeno experiment. The analysis of the standard
experiment is similar. If no distinction is made between the Zeno
and ordinary channels, the decision is based on the number of
absorbed particles $n_a$. If $n_a$ is smaller than or equal to a
decision level $n_a^d$, then $H_1$ is chosen; otherwise $H_2$ is
chosen. The probability of making an error in identifying the gray
level of a given pixel is
\andy{error}
\begin{equation} \label{eq:error}
P_e=\alpha P(H_2|H_1)+(1-\alpha) P(H_1|H_2),
\end{equation}
where
\andy{ph1h2}
\beq \label{eq:ph1h2}
P(H_1|H_2)= \sum_{n_a \leq n_a^d} p(n_a|H_2)
\eeq
is the probability of choosing $H_1$ when $H_2$ is true and
\andy{pnah2}
\beq \label{eq:pnah2}
p(n_a|H_2)=
\left({N\atop n_a}\right) p_a(\tau_2)^{n_a}[1-p_a(\tau_2)]^{N-n_a}
\eeq
is the binomial probability of absorbing $n_a$ particles when
$H_2$ is true. [For $P(H_2|H_1)$ the summation is over
$n_a>n_a^d$]. An optimal protocol is given by determining $n_a^d$
that minimizes the error (\ref{eq:error}).

Alternatively, one defines the likelihood ratio \cite{Helstrom}
\andy{likeli}
\begin{equation}
\label{eq:likeli}
R=\frac{{\cal L}(\tau_1|N,n_a,\alpha)} {{\cal
L}(\tau_2|N,n_a,1-\alpha)},
\end{equation}
where
\andy{probbinomial}
\barr
\label{eq:probbinomial}
\mathcal{L}(\tau_i|N,n_a,\alpha)&=& P_0(H_i) p(n_a|H_i)
\nonumber\\
&=&\alpha\left({N\atop n_a}\right)
p_a(\tau_i)^{n_a}[1-p_a(\tau_i)]^{N-n_a} \;\;\;\;\;\;
\earr
represents the likelihood of hypothesis $H_i$ $(i=1,2)$. The
optimum decision level $n_a^d$ is determined by solving for equal
likelihoods \footnote{The likelihood criterion (\ref{eq:likeli})
is also valid for other (non binomial) statistics and can be
easily generalized to the case of more than two gray levels.}
\andy{R=1}
\begin{equation} \label{eq:R=1}
R=1.
\end{equation}

In both cases one gets
\andy{decbinom}
\begin{equation} \label{eq:decbinom}
n_a^d=\frac{\log\left[\frac{1-\alpha}{\alpha}\right]-
N\log\left[\frac{1-p_a(\tau_1)}{1-p_a(\tau_2)}\right]}
{\log\left[\frac{p_a(\tau_1)}{p_a(\tau_2)}\right]-
\log\left[\frac{1-p_a(\tau_1)}{1-p_a(\tau_2)}\right]}
\end{equation}
and, substituting in (\ref{eq:error}), \andy{beta}
\begin{eqnarray} \label{eq:beta}
P_e&=&\alpha\{1-B_I[N-\tilde{n}_a^d,1+\tilde{n}_a^d;1-p_a(\tau_1)]\}
\nonumber \\ &&+(1-\alpha)
B_I[N-\tilde{n}_a^d,1+\tilde{n}_a^d;1-p_a(\tau_2)],
\end{eqnarray}
where $B_I(a,b;z)$ is the regularized incomplete Beta function
\cite{Abramovitz} and $\tilde{n}_a^d$ is the greatest integer less
than or equal to $n_a^d$. The mean number of absorbed particles is
\andy{mean}
\begin{equation} \label{eq:mean}
N_a^{\mathrm{(Ze)}}=N[\alpha p_a(\tau_1)+(1-\alpha)p_a(\tau_2)].
\end{equation}
By plugging Eqs.\ (\ref{eq:decbinom}) and (\ref{eq:mean}) in
Eq.~(\ref{eq:beta}), the average probability of error
(\ref{eq:beta}) can be expressed as a function of $\alpha$,
$\tau\equiv\tau_1$, $d\tau=\tau_2-\tau_1$ and
$N_a^{\mathrm{(Ze)}}$. The probability of error for the standard
setup is obtained in a completely analogous way.

\begin{figure}
\includegraphics[width=\figwidth]{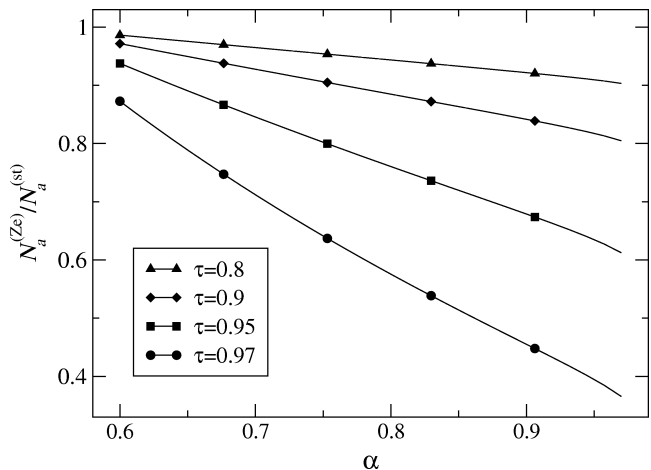}
\caption{\label{fig:absorb} Ratio
of the number of absorbed particles in the Zeno
($N_a^{\mathrm{(Ze)}}$) and standard ($N_a^{\mathrm{(st)}}$)
setup. The smaller the ratio in the graph, the less irradiation in
the Zeno apparatus (for the same resolution). $P_e=0.5\%$;
$d\tau=0.02$; $\tau=\{0.8,0.9,0.95,0.97\}$; $L=2000$.}
\end{figure}

The performances of the Zeno and standard methods are compared in
Fig.\ \ref{fig:absorb}. First, the (exact) $L$th power of the
matrix $V_{\tau}$ in (\ref{eq:Vtau}) is evaluated, then for a
given error rate $P_e$, the number of absorbed particles is
calculated by solving numerically Eq.\ (\ref{eq:beta}). Their
ratio is shown as a function of $\alpha$ for a few values of the
transmission coefficient $\tau$.
Notice that the exposition of the sample can be significantly
reduced if the distribution of gray levels in the sample is not
uniform. For instance, a reduction factor of $2.5$ is obtained
when the sample consists of 97\% of dense material and 3\% of the
less absorbing one, $\alpha=0.97$. Such parameters are typical for
structural analyses: a small structural defect (crack) inside a
thin sample would typically show small contrast ($\delta\tau\ll
1$) with the surrounding almost transparent ($\tau\approx 1$)
material, while its area would be small compared to the area of
the sample ($\alpha\approx 1$).

\subsection{Trinomial (two-channel) protocol}

The binomial decision strategy outlined in the previous subsection
is not the optimum one. Unavoidable losses and other imperfections
of real experimental devices set a strict limit on the maximum
number of loops that can be achieved in a laboratory. In such case
the ordinary channel can no longer be ignored. The data consist
then of the two component vector $(n_z,n_o)$ of the numbers of
particles counted in the Zeno and ordinary output channels. The
decision will be based on \textit{both} these numbers.

\begin{figure}
\includegraphics[width=\figwidth]{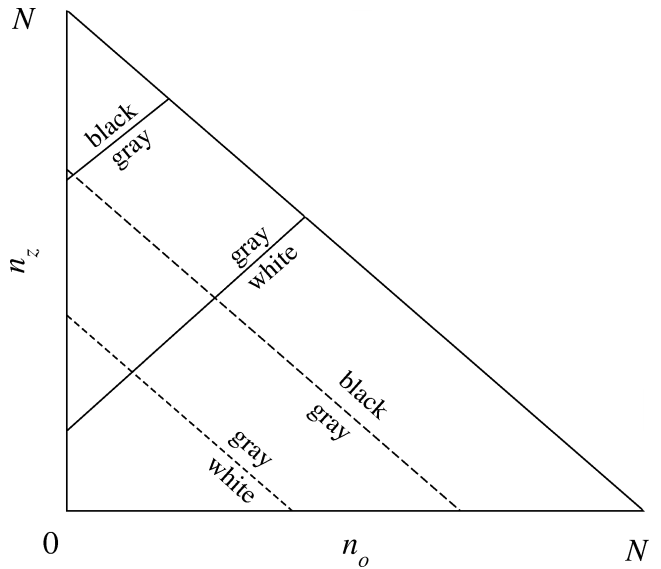}
\caption{\label{fig:levels2} Typical decision levels of the binomial
(dashed lines) and trinomial (solid lines) decision strategies
when $a\approx 1$. This figure corresponds to the simulation shown
in the last row of Fig.~\ref{fig:simul}c, where $M=3$, the
triangle is divided into $M=3$ regions and the three gray shades
are labelled as white, gray and black, respectively.}
\end{figure}

The decision levels are readily obtained from the equal likelihood
criterion
\andy{liktrinomial}
\begin{equation}\label{eq:liktrinomial}
R=\frac{{\cal L}(\tau_1|N,n_z,n_o,\alpha)} {{\cal
L}(\tau_2|N,n_z,n_o,1-\alpha)}=1,
\end{equation}
where
\andy{probtrinomial}
\begin{equation} \label{eq:probtrinomial}
\mathcal{L}(\tau|N,n_z,n_o,\alpha)=
\frac{\alpha N!}{n_z!n_o!n_a!}p_z(\tau)^{n_z}p_o(\tau)^{n_o}
p_a(\tau)^{n_a}.
\end{equation}
This equation is to be solved for the decision vector $(n_z^d,
n_o^d)$. Equation (\ref{eq:liktrinomial}) is just one condition
for the two unknowns $n_z$ and $n_o$, so there exists a
one-parametric family of solutions. By plugging
(\ref{eq:probtrinomial}) into (\ref{eq:liktrinomial}) one easily
finds
\begin{equation}\label{eq:line}\andy{line}
n_z^d-a(\tau_1,\tau_2)n_o^d=b(\tau_1,\tau_2,\alpha),
\end{equation}
where the coefficients $a$ and $b$ read
\andy{coefa}
\andy{coefb}
\begin{eqnarray}
\label{eq:coefa}
a&=&\frac{\log\left[
\frac{p_o(\tau_1) p_a(\tau_2)}{p_o(\tau_2) p_a(\tau_1)}\right]}
{\log\left[
\frac{p_z(\tau_2) p_a(\tau_1)}{p_z(\tau_1)p_a(\tau_2)}\right]},\\
\label{eq:coefb}
b&=&\frac{N\log\left[\frac{p_a(\tau_1)}{p_a(\tau_2)}\right]
+\log\left[\frac{\alpha}{1-\alpha}\right]}
{\log\left[
\frac{p_z(\tau_2) p_a(\tau_1)}{p_z(\tau_1)p_a(\tau_2)}\right]}.
\end{eqnarray}
In a two-dimensional representation each possible experimental
outcome $(n_z,n_o)$ is represented by a point lying inside the
triangle $\{0\le n_o+n_z\le N\}$ shown in Fig.~\ref{fig:levels2}.
Equation (\ref{eq:line}) divides this triangle into two regions.
All experimental outcomes that fall within the same region issue
the same decision. In the general case of $M$ different gray
levels there are $M-1$ equations (\ref{eq:line}) defining $M-1$ in
general non-parallel lines dividing the square into $M$ strip-like
regions. This is shown in Fig.\ \ref{fig:levels2} for $M=3$.

An interesting situation arises when the coefficient $a$ in
(\ref{eq:line}) becomes close to unity. In that case, the decision
level is the line $n^d_z-n^d_o=\mathrm{const.}$ Let us recall that
the decision levels of the binomial decision strategy discussed in
the previous subsection were $n^d_a=\mathrm{const.}$ (see
(\ref{eq:decbinom})), or, equivalently,
$n^d_z+n^d_o=\mathrm{const.}$ Hence if $L$, $\tau_1$ and $\tau_2$
are such that $a\approx 1$, the decision levels of the binomial
and trinomial decision strategies are \textit{orthogonal} to each
other. This is shown in Fig.~\ref{fig:levels2}. Under such
conditions one can expect further gain in the precision of the
Zeno apparatus as compared to standard absorption tomography. This
regime was chosen for our computer simulations of the following
section.

Notice that the steepness of the decision lines (\ref{eq:line})
depends \textit{only} on the absorption of the corresponding
adjacent gray levels. It depends \textit{neither} on their
frequencies, \textit{nor} on the total number of incident
particles.

Finally, let us discuss the limit $L\rightarrow \infty$. When the
number of loops $L$ increases, $p_o\rightarrow 0$ much faster than
$p_a$ [see (\ref{eq:pzoa})] and one is allowed to put $n_o^d=0$
and $n_z^d=N-n_a^d$ in Eq.~(\ref{eq:line}), which reduces to
\andy{nda}
\beq
\label{eq:nda}
n_a^d=N-b(\tau_1,\tau_2,\alpha).
\eeq
By substituting $p_z(\tau)=1-p_a(\tau)$ in
(\ref{eq:coefb})-(\ref{eq:nda}), one reobtains the binomial
condition (\ref{eq:decbinom}): the binomial strategy becomes
optimal in this limit.

\section{Simulations}

\begin{figure}
\includegraphics[width=0.5\columnwidth]{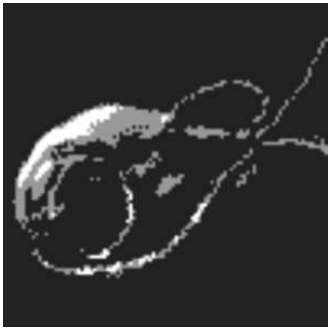}
\caption{The object to be reconstructed: a cell of \textit{Giardia lamblia},
one of the most primitive eukaryotes. The original picture has
been reduced for simplicity to three levels of gray: white, gray
and black, occurring with frequencies $\alpha_w=0.02$,
$\alpha_g=0.07$ and $\alpha_b=0.93$ respectively.}
\label{fig:object}
\end{figure}
\begin{figure*}
\includegraphics[width=\textwidth]{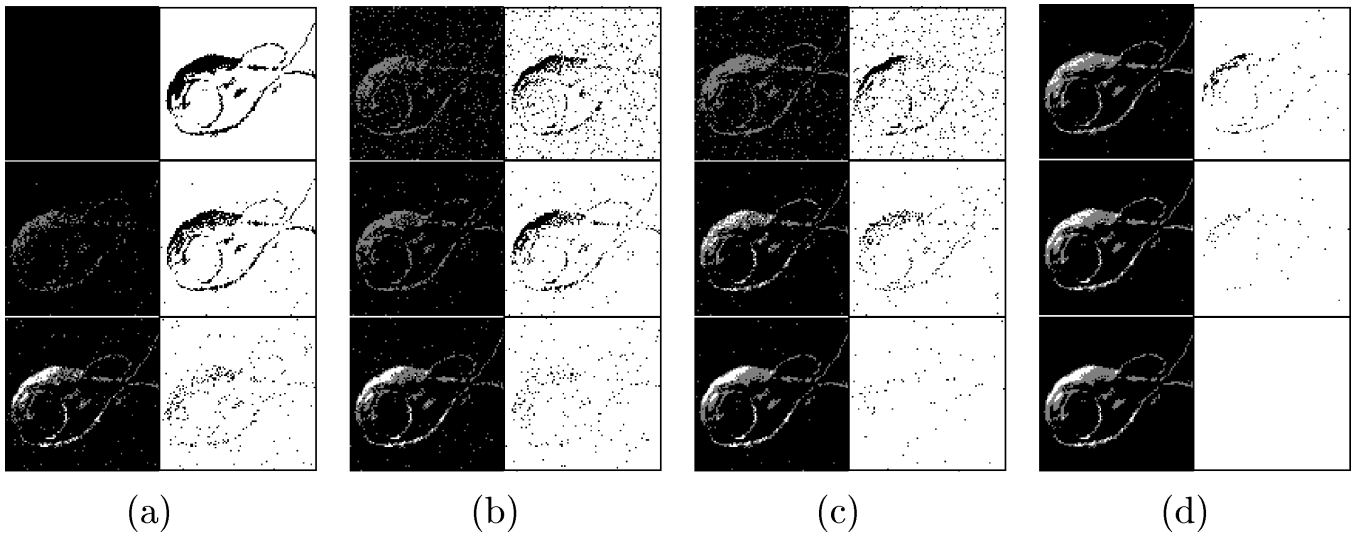}
\caption{\label{fig:simul} Comparison of standard and Zeno
tomographic techniques. In each frame: top left=reconstruction by
standard technique; top right=misinterpreted pixels by the
standard technique; center left=reconstruction by Zeno technique
with $L=10$; center right=misinterpreted pixels by the Zeno
technique with $L=10$; bottom left=reconstruction by Zeno
technique with $L=165$; bottom right=misinterpreted pixels by the
Zeno technique with $L=165$. The mean number of absorbed particles
per pixel (irradiation) is $N_a=1.7$, $2.3$, $4$ and $13$ for
frames (a), (b), (c) and (d) respectively. The total number of
particles $N$ (total energy) scales approximately like $3N_a$,
$1.8N_a$, and $6.5N_a$ for top, center and bottom reconstructions,
respectively. We used $\tau_w=0.99$, $\tau_g=0.96$, and
$\tau_b=0.8$. The sample consists of 10,000 ($=100
\times 100$) pixels, where white, gray and black occur with
frequencies $\alpha_w=0.02$, $\alpha_g=0.07$ and $\alpha_b=0.93$,
respectively. The number of misinterpreted pixels are (top to
bottom): (a) $968$, $786$, $315$; (b) $942$, $596$, $212$; (c)
$717$, $382$, $68$; (d) $205$, $69$, $0$. }
\end{figure*}

We have seen that the Zeno technique can reduce the level of
absorption without losing resolution (compared to the standard
technique). (Alternatively, the Zeno setup can yield an improved
resolution, while keeping the absorption at the same level of the
standard setup.) The object in Fig.\ \ref{fig:object} is a cell of
\textit{giardia lamblia}, a protist, one of the most primitive
eukaryotes. Giardia has been called a ``missing link" in the
evolution of eukaryotic cells from prokaryotic cells. The number
of gray levels in the figure has been reduced to three to make the
analysis simpler: white, gray and black, $\tau_w=0.99$,
$\tau_g=0.96$, $\tau_b=0.8$, occurring with frequencies
$\alpha_w=0.02$, $\alpha_g=0.07$, and $\alpha_b=0.93$
respectively. Figure
\ref{fig:simul} shows the results of a numerical simulation,
performed with the standard and Zeno methods, the latter for
$L=10$ and $L=165$, for different numbers of absorbed particles
$N_a$. In each frame the standard and the two Zeno reconstructions
are compared, together with the pixels that have been
misinterpreted. Figure
\ref{fig:simul} confirms the expectation based on the asymptotic
formulas (\ref{eq:pzoa}): in general, provided that the object
contains a small fraction of more transparent pixels and a larger
fraction of more absorbing material, the Zeno setup yields a
better resolution for a given irradiation. Clearly, a significant
improvement with respect to standard absorption tomography is
achieved for as few as $L=10$ loops. The improvement is very large
for $L=165$.

The number of absorbed particles increases from (a) to (d) in
Fig.\ \ref{fig:simul}. Observe that in (a) the standard
reconstruction fails completely, while the outline and basic shape
of the object can be recognized already in the Zeno reconstruction
with $L=10$. In (c) the Zeno reconstructions are quite good, while
standard tomography does not detect white pixels in the object.
When the intensity of the illuminating beam is increased further,
in frame (d), all the reconstructed images become visually hard to
tell from the sample, but the error rates of the Zeno apparatuses
are still much better (by a factor three or more), as shown by the
number of misinterpreted pixels.

It is worth commenting on the distribution of misinterpreted
pixels. Clearly, in all the cases analyzed, it is not uniform. In
general, when the distribution of gray levels in the sample is not
uniform, any reconstruction technique tends to perform better in
the ``background," while making more mistakes in the region where
the ``structure" is present. The improvement due to the Zeno
method becomes apparent if one looks in particular at Figures
\ref{fig:simul}(b) and (c): in these cases, interestingly, the
standard method yields more mistakes in the background; this is an
unpleasant feature, if one is interested in detecting small
irregular structures in a more or less uniform background. The
features of the distribution of misinterpreted pixels require more
careful study and their comprehension might lead to additional
ideas.

Any increase in the number of loops $L$ in the interferometer
makes the difference between standard and Zeno tomography even
bigger. Clearly, this is more demanding in terms of experimental
realization.

\section{Conclusions}
\label{sec-conc}\andy{sec-conc}
We have shown that a quantum Zeno tomography is possible and
performs better than standard tomography if a given prior
knowledge about the distribution of grays in the sample is
available. This is a common situation in radiography, where one is
often interested in detecting a small structure in a uniform
background, like for instance in the analysis of small structural
defects.

In our numerical simulations we have illustrated some situations
in which the resolution is improved by the Zeno method, for a
\textit{given number of absorbed particles}. Alternatively, for a
\textit{given resolution}, the Zeno method performs better, absorbing
less particles. This can be interesting in applications, for
instance if one wants to limit the damage provoked by the
absorption of radiation without losing in resolution.

It is obvious from Fig.\ \ref{fig:absorb} that an even larger
improvement is possible for almost transparent samples, provided
that $\alpha$ is close to unity. This means that there is
\textit{no fundamental limit} on the improvement that can be
achieved over the standard setup: in other words, there is no
``optimal" configuration.

There are additional issues that deserve careful study. For
instance, the effects due to a Poissonian beam (total number of
incoming particles $N$ not fixed) and a complex transmission
coefficient $\tau$.

Let us also comment on experimental feasibility. Figure
\ref{fig:simul} shows that an experimental test of the
Zeno tomographic technique should not be as difficult as one might
think: simulations have been performed for as few as $L=10$ loops
in the interferometer, giving better results than the standard
method. It is reasonable to think that a Zeno setup with a much
larger number of loops can be built for UV light (highly absorbed
by some biological samples). Also, by changing the light
wavelength, one could efficiently ``observe" different regions of
the sample (or slightly different samples). Moreover, the
experimental configuration we have proposed (photons in a MZI,
like in Fig.\ \ref{fig:outline}) is certainly not the only
conceivable one. Phase imaging and tomography have been
demonstrated for both X-rays and neutrons
\cite{Bonse}. More to this, Rauch and collaborators, with the
VESTA apparatus
\cite{VESTA}, have been able to keep \textit{neutrons} in a
one-meter long perfect crystal storage system (``resonator") for a
few seconds, so that the neutrons bounce back and forth between
two mirrors \textit{several thousands} times. This would lead us
to the full asymptotic ($L\gg 1$) regime considered in
Fig.~\ref{fig:absorb} and last row of Fig.\ \ref{fig:simul}, where
the Zeno method  can perform much better. There is hopefully more
to come.

\begin{acknowledgments}
This paper is dedicated to Professor Jan Pe\v{r}ina on the
occasion of his 65th birthday. This work is supported by the
TMR-Network of the European Union ``Perfect Crystal Neutron
Optics" ERB-FMRX-CT96-0057. J.\ \v{R}.\  and Z.\ H.\ acknowledge
support by the project LN00A015.
\end{acknowledgments}


\end{document}